\documentstyle[graphicx]{mn2e}
\footnotesize
\newdimen\digitwidth    
\setbox0=\hbox{\rm0}
\digitwidth=\wd0
\catcode`!=\active
\def!{\kern\digitwidth}
\normalsize

\title{Observations of radio pulses from CU Virginis}
\author[V. Ravi et al.]{V. Ravi,$^{1,2,3}$\thanks{Email: v.vikram.ravi@gmail.com}
G. Hobbs,$^2$ 
D. Wickramasinghe$^{4}$, 
D. J. Champion,$^{2,5}$ 
M. Keith,$^2$ 
\newauthor
R. N. Manchester,$^2$
R. P. Norris,$^2$ 
J. D. Bray,$^{6,2}$ 
L. Ferrario,$^4$ 
D. Melrose$^7$
\\
$^1$ Research School of Astronomy and Astrophysics, ANU, Mt. Stromlo Observatory, 
Weston ACT 2611, Australia \\
$^2$ Australia Telescope National Facility, CSIRO Astronomy and Space Sciences, PO~Box~76, Epping NSW~1710, Australia\\
$^3$ Space Sciences Laboratory, University of California, Berkeley, CA 94720, USA\\
$^4$ Mathematical Sciences Institute, The Australian National University, Canberra, 
ACT 0200, Australia \\
$^5$ Max-Planck-Institut f\"{u}r Radioastronomie, Auf dem H\"{u}gel 69, 53121, Bonn, 
Germany \\
$^6$ School of Chemistry \& Physics, The University of Adelaide, SA 5005, Australia \\
$^7$ School of Physics, The University of Sydney, NSW 2006, Australia \\
} 
\date{}

\pagerange{\pageref{firstpage}--\pageref{lastpage}}
\pubyear{2010}

\begin{document}

\maketitle

\label{firstpage}

\begin{abstract}
The magnetic chemically peculiar star CU Virginis is a 
unique astrophysical laboratory for
stellar magnetospheres and coherent emission processes. It is the only known
main sequence star to emit a radio 
pulse every rotation period. Here we report on new observations of the CU
Virginis pulse profile in the 13 and 20\,cm radio bands. 
The profile is known to be characterised by two peaks of 100$\%$
circularly polarised emission that are thought to arise in an 
electron-cyclotron maser mechanism.  We find that the trailing peak is 
stable at both 13 and 20\,cm, whereas the leading peak is intermittent at 13\,cm.
Our measured pulse arrival times confirm the discrepancy previously 
reported between the putative stellar rotation rates
measured with optical data and with radio observations. We suggest that this period 
discrepancy might be caused by an unknown companion or by instabilities in the 
emission region. Regular long-term pulse 
timing and simultaneous multi-wavelength observations are essential to clarify 
the behaviour of this emerging class of transient radio source.
\end{abstract}


\section{Introduction}

CU Virginis (HD124224, hereafter CU Vir) is unusual in that it
is a stable source of coherent, polarised radio emission. Observations in
the 13, 20 and 50\,cm radio bands have revealed one or more short
duty-cycle emission peaks per rotation period (0.5207 days) that
re-appear at the same rotation phases \cite{tll+00,tlu+08,gs10}.  These
peaks are distinguished from the quiescent emission also observed from 
the source by their 100$\%$ circular polarisation, short
durations ($<$1 hour) and flux densities of up to a factor of six above
the quiescent levels. Together, we refer to the peaks as the CU Vir pulse.

As a magnetic chemically peculiar A0 (sometimes B9) star, CU Vir is 
thought to possess an offset dipole magnetic field \cite{d+52,h+97}, with a 
pole strength of $\sim$3\,kG \cite{tll+00}, misaligned from the rotation axis 
\cite{bl+80}. Various stellar parameters are summarised in Table 1, with errors 
in the last decimal places given in parantheses. The pulses are thought
to be emitted in the vicinity of one magnetic pole \cite{tll+00}. The
pulse emission geometry is strikingly similar to the
canonical model for radio pulsars (e.g. Manchester \& Taylor
1977)\nocite{mt+77}. The beaming of the CU Vir pulsed emission, the 
100$\%$ circular polarisation of the pulse, the high brightness
temperature of $>$10$^{12}$\,K and the flat spectrum between 13 and
20\,cm are all indicative of coherent emission from an
electron-cyclotron maser (ECM) mechanism \cite{tll+00,kgb+07,tlu+08}. ECM
emission is characterised by narrow bandwidths corresponding
to either the fundamental or any of the first few harmonics of the local cyclotron
frequency,
\begin{equation}
\nu_{B}=\frac{q_{e}B}{2\pi m_{e}},
\end{equation}
where $q_{e}$ and $m_{e}$ are the elementary charge and electron mass
respectively, and $B$ is the magnetic field strength
\cite{md+82}. 

\nocite{s+98}
\begin{table}
\begin{center}
\caption{Stellar parameters of CU Vir from Trigilio et al. (2000) and SIMBAD.}
\label{table1}
\begin{tabular}{ll}
\hline
Property & Value \\
\hline 
Position (J2000) & 14:12:15.80, +02:24:34.0 \\
Spectral type & A0p ($\alpha^{2}$ CVn) \\
Distance & 80(6) pc \\
Rotation period & 0.5207 days \\
Stellar radius & 2.2(2) solar radii\\
Magnetic field strength & 3.0(2)$\times10^{3}$ G \\
Mass (1) & $\sim$3 solar masses \\
\hline
\end{tabular} 
\end{center}
\medskip
(1) This mass estimate is from St{\c e}pie{\'n} (1998).
\end{table}

By comparing 20\,cm observations of two pulses in 1999 
with a pulse recorded in 1998, Trigilio et al. (2008) measured a 
radio period that was 1.2\,s slower than the latest optical rotation 
period \cite{prm+98}. The Pyper et al. (1998) rotation 
ephemeris for CU Vir was derived using data extending up to 1997, 
and no new optical rotation ephemeris has since been published. Pyper et 
al. (1998) also found that the optical period itself had reduced by 
$\sim$2\,s in 1998.  These periodicity `glitches' are not well understood.

Here we present new observations of CU Vir pulse profiles in the 13
and 20\,cm bands. In \S2 we detail the observations and data
reduction. We describe our results, for both our data and for archival
data, in \S3, and confirm the radio period discrepancy reported by
Trigilio et al. (2008).  In \S4 we discuss implications for the CU Vir magnetic field, 
and, in \S5, we outline various hypotheses for the period discrepancy. We 
present our conclusions and discuss future work in \S6.

\section{Observations and data analysis}

We observed CU Vir on 2008 October 31 with the Australia Telescope
Compact Array (ATCA).  The array of six 22-metre dishes, in its
fully-extended 6A configuration\footnote{http://www.narrabri.atnf.csiro.au/observing/}, recorded data in 32 channels
across 128\,MHz bandwidths centred at 1.384\,GHz (20\,cm) and
2.638\,GHz (13\,cm). Complex cross-correlations (visibilities), integrated over
10\,s intervals, were recorded for all baselines in all Stokes parameters. 
The absolute flux
density scale and the frequency response over the receiver bandpasses
were characterised using the radio galaxy PKS B1934$-$638. 
The gains of the individual antennas, atmospheric and
instrumental path-length variations for each antenna and signal path,
and the cross-talk between the orthogonal linear feeds in each antenna
were calibrated using five-minute observations of the radio galaxy PKS
B1416$-$067 at 20-minute intervals during the observations. The total
observing time on CU Vir was 7.75 hours. We pointed the antennas 10\arcsec~South 
of CU Vir to avoid source contamination by `non-closing' correlator offsets.

We reduced our data using the MIRIAD set of software routines \cite{stw95}. 
In order to optimise our source-subtraction technique, the four shortest
baselines in each configuration were removed from the
data, along with radio-frequency interference. Multi-frequency synthesis 
images with extent four primary beam full-width half-maxima 
were made of the source
field in the 13 and 20\,cm bands. Attenuation caused by the spatial response of 
the primary beam was corrected for. All sources detected at greater than
five standard deviations of the noise in the images were subtracted. Our
subtraction technique involved producing CLEAN models of each source
and subtracting them from the visibility data using the MIRIAD task
UVMODEL. We then shifted the phase centre of the visibilities to the
CU Vir position, and inspected the time-series of the real components
of the visibilities, averaged over two-minute intervals, for Stokes V
pulses.  Having identified the pulse durations, we imaged
the off-pulse emission, produced CLEAN models in the 13 and 20\,cm bands 
and subtracted them from the visibility data.

Lightcurves of the pulsed emission were then produced in Stokes I
and V from the source-subtracted visibility data by averaging the
real visibility components over two-minute intervals.  Stokes I
lightcurves of the off-pulse emission were also produced by averaging
the real visibility components of the off-pulse datasets over 50
minute intervals and the errors were scaled appropriately. 

We repeated our data reduction procedure for the CU Vir datasets
(obtained from the Australia Telescope Online 
Archive\footnote{http://atoa.atnf.csiro.au}) described in
Trigilio et al. (2008).  In summary, we analysed observations of CU
Vir from three epochs: Epoch A (1999 May 29), Epoch B (1999 August 29)
and Epoch C (2008 October 31). Data from Epochs A and B were collected
by Trigilio et al. (2008), and the Epoch C observations were
ours. Details of all observations are summarised in
Table~\ref{table2}.

\begin{table}
\begin{center}
\caption{Basic parameters and detections from the three datasets
analysed. \textit{Q}: quiescent
detections. \textit{P}: Number of peaks detected.  
The Stokes I quiescent ($Q_{20,13}$) and the Stokes V peak ($P_{20,13}$) 
flux densities (in mJy) for all peaks detected are also given. The errors in the peak flux densities are approximately 
1\,mJy. }
\label{table2}
\begin{tabular}{lllll}
\hline
 & Epoch A & Epoch B & Epoch C \\
\hline
Date (UT) & 29/05/99 & 29/08/99 & 31/10/08 \\
Timespan (hrs) & 9.3 & 9 & 9 \\
ATCA config. & 6A & 6D & 6A \\
20 cm $Q$ & Yes & Yes & Yes \\
13 cm $Q$ & Yes & Yes & Yes \\
20 cm $P$ & 2 & 2 & 2 \\
13 cm $P$ & 1 & 1 & 2 \\
\hline 
$Q_{20}$ & $3.4(3)$ & $2.3(2)$ & $2.5(3)$ \\
$P_{20}$ & 5, 13 & 3, 6 & 11, 12 \\
$Q_{13}$ & $3.4(2)$ & $3.0(2)$ & $2.9(2)$\\
$P_{13}$ & 0, 14 & 0, 6 & 12, 8 \\
\hline
\end{tabular} 
\end{center}
\end{table}

\section{Results}

\begin{figure}
\centering
\includegraphics[height=8cm,angle=-90]{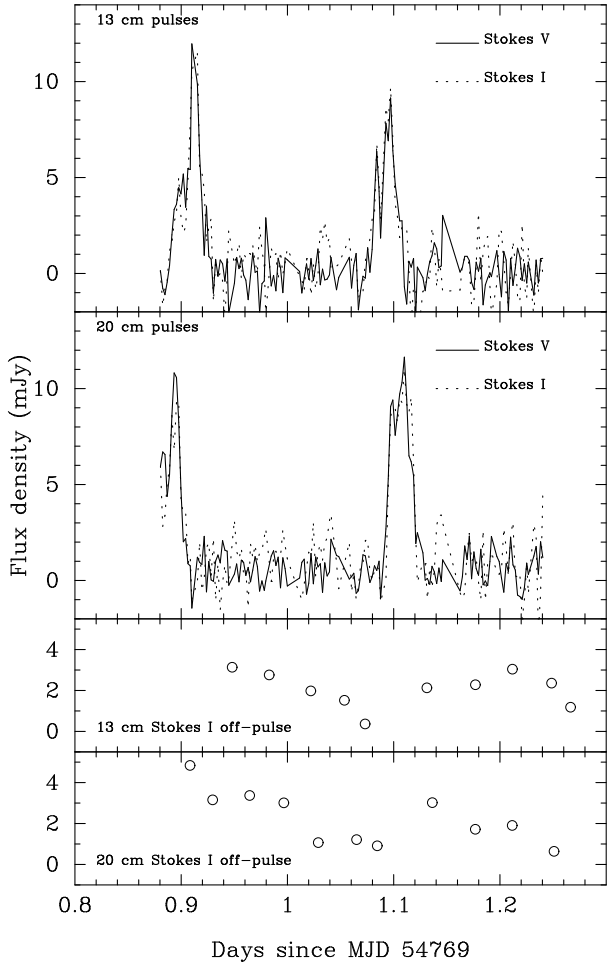}
\caption{Epoch C pulses and quiescence from CU Vir. From top to bottom: 
13\,cm Stokes I (dotted) and V pulses, 20\,cm Stokes I (dotted) and V pulses, 
13\,cm Stokes I off-pulse flux densities and 20\,cm Stokes I off-pulse flux densities. 
The pulse lightcurves are averaged in two-minute bins, and the off-pulse lightcurves 
in 50-minute bins. The off-pulse measurements have errors of 0.4\,mJy. }
\end{figure} 

We observed two peaks of 100$\%$ right-circularly polarised emission
from CU Vir during Epoch C at both 13 and 20\,cm, as well as
time-variable quiescent emission. The pulse and quiescent lightcurves
are shown in Figure 1. No significant linear polarisation was detected
in the pulses, and the quiescent emission was found not to be
significantly polarised. Our analysis method reproduces the results of
Trigilio et al. (2008) for Epochs A and B.  In Figure 2 we plot the
Stokes V lightcurves for all three epochs aligned according to the
rotation ephemeris of Pyper et al. (1998). The average 
quiescent flux densities are presented in Table~\ref{table2} along with 
the peak flux densities for all epochs.

Our Epoch C observations clearly show two peaks 
at both 13 and 20\,cm, in contrast to the lone 13\,cm peaks in the 
Epoch A and B observations (see Figure 2). The Epoch C peak separations are smaller in the
13\,cm band (separation of $\sim$4\,hr) than in the 20\,cm band
(separation of $\sim 5$\,hr), but the midpoints between the peaks in both bands
occur at the same time. This implies that, if pulse 
arrival times at different frequencies need to be
compared, the ``arrival-time'' of the pulse should be taken as the
midpoints between the peaks.  

Figure~2 clearly shows that the rotational ephemeris of Pyper et
al. (1998) is not able to align the observed pulses.  In order to
obtain an accurate ephemeris for CU Vir we obtained pulse
times-of-arrival (TOAs), as described above, and used standard pulsar timing software
\cite{hem+06} to fit simple periodicity models to these TOAs 
using uniform weighting. The
resulting best-fit period was
$44989.967(8)$\,s$=0.52071721(9)\times10^{-8}$\,days. The post-fit 
rms timing residual was 46\,s. No significant period derivative could be 
determined. Our measured period is 1.221(8) seconds slower than that of 
Pyper et al. (1998) which is consistent with that determined by 
Trigilio et al. (2008) using pulses separated by $\sim$1\,yr.

\begin{figure}
\centering
\includegraphics[height=7cm,angle=-90]{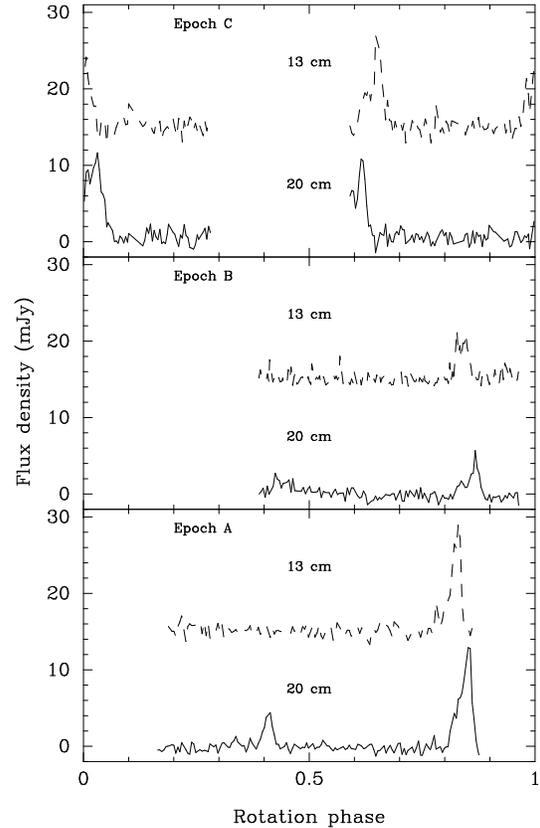}
\caption{Illustration of the varying pulse phases with respect to the 
ephemeris of Pyper et al. (1998). The pulse profiles from Epochs A, B and C are 
shown in the bottom, middle and top panels respectively. 
The 13\,cm Stokes V pulses (dashed lines) are placed at +15\,mJy to 
differentiate them from the 20\,cm Stokes V pulses (solid lines).}
\end{figure}

\begin{table}
\begin{center}
\caption{List of TOAs with references derived from all lightcurves of CU Vir.}
\label{table5}
\begin{tabular}{llll}
\hline
TOA (MJD) & $\nu_{obs}$ & Epoch & Reference \\
         & (GHz) & \\
\hline
50966.159(8) & 1.425  & June 98 & Trigilio et al. (2000) \\
51327.535(3) & 1.380 & A & Trigilio et al. (2008) \\
51419.189(3) & 1.380 & B & Trigilio et al. (2008) \\
54770.007(3) & 1.380 & C & this paper \\
54770.007(3) & 2.368 & C & this paper \\
\hline
\end{tabular} 
\end{center}
\end{table}

\section{The magnetic field of CU Vir}

Spectrally peculiar A and B stars (Ap/Bp stars), 
such as CU Vir, have long been known to be highly magnetic \cite{b47}, commonly 
with large dipolar components (e.g. Borra $\&$ Landstreet 1980). A leading hypothesis for the 
field origins is that of `frozen-in' fossil fields from early evolutionary stages \cite{dl09}. 
The frozen-in field paradigm, numerically simulated by Braithwaite $\&$ Nordlund 
(2006)\nocite{bn06}, involves an unstable protostellar magnetic field, concentrated 
in the core, that quickly stabilises over a few Alfv{\`e}n timescales, given by 
$\tau_{A}=\frac{R_{*}}{v_{A}}$, 
where $R_{*}$ is the stellar radius and $v_{A}$ is the Alfv{\`e}n speed. 
For Ap stars such as CU Vir, $\tau_{A}\sim10$\,yr. For most of the main-sequence life 
of the star, the field then slowly expands outwards through ohmic diffusion, creating 
a slowly-increasing atmospheric poloidal field component.

Simultaneous observations of the CU Vir pulse at many radio frequencies could provide 
interesting insights into the magnetic field structure. ECM emission frequencies directly map 
to magnetic field strengths (see Equation 1), and the emission is tightly beamed at 
fixed angles to field lines determined by the kinematics of the radiating electrons 
\cite{md+82}. Our dual-frequency observations of the 
CU Vir pulse show a time difference of $\sim$30 minutes 
between correponding peaks. This translates to a difference in the angle of emission with 
respect to the magnetic axis of 14$^{\circ}$. Assuming ECM emission at the 
fundamental cyclotron frequency, the 13\,cm and 20\,cm bands correspond to emission from 
magnetic field regions of 850\,G and 500\,G respectively. 
If the pulse is emitted from field lines occupying a narrow range of magnetic co-latitudes \cite{ltb+06,tlu+08}, 
the tangents to the field lines at field strengths of 850\,G and 500\,G regions 
must also differ in angle to the magnetic axis by 14$^{\circ}$. Assuming a dipole field for 
CU Vir, this occurs for field lines that intersect the magnetic equator at 
radii of $\sim1.5R_{*}$. This is much closer than the 12-17$R_{*}$ predicted in the 
magnetospheric model of Leto et al. (2006) for field lines along which 
the radiating electrons propagate. Hence, we suggest that either the magnetic field is more curved than for a pure dipole in the 
polar regions, or that the Leto et al. 
(2006) model for the CU Vir magnetosphere does not adequately account for the CU Vir pulses. 
We are conducting wide-band radio studies of the CU Vir pulse 
to further probe the magnetic field structure in the pulse emission regions.

\section{The anomalous radio periodicity}

The offset between the radio pulse- and optical variability-periods of CU Vir 
can be interpreted in a variety of ways. 
Pyper et al. (1998) reported a reduction in the optical period of
CU Vir of $\sim2$\,s that occurred in 1984. This
was derived by fitting different periods to time-resolved
spectrophotometric data gathered before and after 1984. A possible
interpretation of the period discrepancy between the radio pulse data
taken after 1998 and the latest ephemeris published by Pyper et
al. (1998) is that the CU Vir rotation period has again
decreased, sometime between the epoch of the last data analysed by
Pyper et al. (1998) (May 1997) and the Epoch A radio observations.

Trigilio et al. (2008) suggested that the loss of all the confined 
magnetospheric mass, as modelled by Havnes $\&$ Goertz (1984)\nocite{hg84}, at the 
epochs of the period changes could explain the apparent 0.2\,Myr spin-down timescale. 
Besides CU Vir, some Ap/Bp stars have been observed to have steadily reducing 
rotation periods \cite{toc+10}, but with spin-down timescales of $\sim$1\,Myr. 
Such spin-down timescales are predicted by simulations of steady 
angular momentum loss from Ap/Bp stars to the magnetic field and 
magnetically-confined wind, with episodes of confinement-breaking during which the 
magnetosphere is emptied \cite{uot09}. In contrast to the suggestion of Trigilio et al. (2008), 
the emptying episodes are associated with \textit{less} angular momentum loss, as 
the star can no longer lose angular momentum to its surroundings. These simulations 
cannot reproduce the apparent spin-down of CU Vir. We therefore explore other 
possibilities for the CU Vir period discrepancy.

\subsection{A drifting pulse emission region}

If we assume that the CU Vir rotation period has \emph{not} changed,
then two possibilities exist: either the pulse periodicity we measure
is real and represents a stable, persistent effect, or the periodicity
is coincidental and future radio pulse measurements will not follow
our fitted pulse period. We first consider the former case, and
the latter case in \S\ref{sec:unstable}.

Our measured period discrepancy could imply a non-fixed emission
region that steadily drifts in azimuth about the rotation axis. The rate of
drift corresponds to the period difference, $\sim$1.2\,s, per
optical period. Thus, the drift period is approximately 53 years. 
This drift could be caused by a mechanism that is 
analogous to the differential rotation in the 
solar magnetosphere \cite{s89}.

The emission region might also be coupled to the 
orbit of an object with a 53 year orbital period. This would cause the emission region to 
drift about the orbital axis. Whatever the relative orientation of 
the orbital angular momentum axis and CU Vir rotation axis may be, a systematic change with time 
in the pulse profile, both in shape and frequency characteristics, is expected 
because the emission region would be drifting relative to the magnetic field. 
Keplerian dynamics place such an object at a radius of approximately 
20\,AU. Such a system could be directly analogous to the Jupiter-Io interaction, 
where the orbital phase of Io around Jupiter is strongly correlated with the 
occurrence of Jovian decametric emission \cite{b64}.

\subsection{An unstable emission region}\label{sec:unstable}

The discrepant radio pulse periodicity could be ascribed to
coincidence, and could indicative of an unstable emission
region. In this interpretation, future measurements of the radio
lightcurve will not fit the currently measured pulse
period. 

Pulse shape variability is also prevalent among the only other stable emitters of 
coherent radio pulses: radio pulsars. Indeed, single pulses from pulsars vary
greatly in both structure and power from pulse to pulse. It has
however been shown that pulsars have extremely stable characteristic
pulse profiles, formed by averaging large numbers of individual pulses
together \cite{mt+77}. A similar average pulse
profile for CU Vir, attempted by Kellett et al. (2007)\nocite{kgb+07}, will be useful in 
ascertaining the emission region structure and stability, as well as 
aiding in pulse timing.

\section{Conclusions and future work}

For more than a decade, CU Vir has been known to be unique among main-sequence 
stars in producing strong, periodic peaks of coherent radio emission. Our 
observations reveal, for the first time, twin peaks in the CU Vir pulse profile at both 13 
and 20\,cm. While the 13\,cm peak separation is one hour 
less than the 20\,cm peak separation, the midpoints of the peaks occur 
simultaneously in both bands. We show that the arrival-time difference between the 
13\,cm and 20\,cm peaks could indicate that the field structure is more complex than a 
pure dipole. We demonstrate that a characteristic pulse arrival 
time can be determined from the midpoint of the peaks in the profile. 
Using four arrival times derived with this technique from archival data as 
well as our own observations, we find that the radio pulses 
fit a periodicity that is 1.221\,s slower than the most recently 
published optical rotation period. This 
confirms, over a 10 year period, the initial trend evinced by Trigilio et al. (2008) 
using pulses separated by one year. We suggest that, in contrast to the 
explanations of Trigilio et al. (2008) for the period discrepancy, the anomalous radio 
periodicity could be caused by a drifting or unstable emission region.

Targeted observations can reveal the cause of the period discrepancy. We plan a simultaneous 
measurement of a radio pulse and the optical lightcurve of CU Vir to determine 
whether a period change has indeed occurred. If a mass-loss event is the cause of the 
period change, a detached mass shell could be directly detected and its shape measured 
using infrared or optical interferometry. Thermal emission from electron-hydrogen 
atom collisions could be expected from a cooling mass shell \cite{tt+06}. Hydrogen 
recombination lines could also be a significant emission component, particularly at 
optical and near-infrared wavelengths. Furthermore, if a mass-loss event was 
associated with the 1984 optical period reduction, 
a much bigger shell might also be visible. If the pulse emission region is 
coupled to the orbit of a companion, 
sensitive radio very long baseline interferometry could detect 
radio emission from plasma flows between the companion and CU Vir, as well as possible 
reflex motion of CU Vir. Conventional planet detection techniques \cite{j+09}, 
such as time-resolved optical photometry and radial velocity measurements, are not 
applicable to CU Vir given its large degree of intrinsic variability. However, 
long-period binarity could be indicated by variations in the timing of the extrema of 
the optical lightcurve caused by gravitational effects in the binary.

Despite the many remarkable properties of CU Vir, including its close proximity 
to the Earth and its fast rotation rate, it is unlikely that it will remain unique as 
a source of radio pulses. Future large-area continuum 
surveys by next-generation radio telescopes, such as the Australian Square Kilometre Array 
Pathfinder, the Karoo Array Telescope and eventually the Square Kilometre Array 
will potentially find many similar objects, leading to further insights into this 
fascinating class of radio transient.

\section*{Acknowledgements}

We thank the referee, Ian Stevens, for many valuable suggestions. 
We are also grateful for the advice and insight of C. Trigilio on this letter. 
The Australia Telescope Compact Array is part of the Australia
Telescope which is funded by the Commonwealth of Australia for
operation as a National Facility managed by CSIRO. This research has 
made use of the SIMBAD database, operated at CDS, Strasbourg, France. 
GH is supported by an Australian Research Council
QEII Fellowship (project \#DP0878388).

\bibliography{vikram}
\bibliographystyle{mn}

\end{document}